\documentclass[a4paper,12pt]{article}
\pdfoutput=1 

\usepackage{jheppub}
\usepackage{hyperref} 

\usepackage[T1]{fontenc} 
\usepackage{graphicx}
\graphicspath{figures/}

\title{\boldmath Resonant leptogenesis in minimal inverse seesaw ISS(2,2) model}

\author{Bikash Thapa}
\author{and Ng. K. Francis}

\affiliation{Department of Physics, Tezpur University, Tezpur - 784028, India}

\emailAdd{bikash2@tezu.ernet.in}
\emailAdd{francis@tezu.ernet.in}

\abstract{We investigate the parameter space of the minimal inverse seesaw ISS(2,2) model for successful leptogenesis. The framework of ISS(2, 2) is realized by augmenting the Standard Model with two right-handed and two Standard Model singlet neutrinos. The decay of the heavy sterile states which is essentially an admixture of the right-handed and SM singlet neutrino states produces the baryon asymmetry of the universe. In this predictive model of leptogenesis, we study resonant leptogenesis where the mass splitting between the heavy sterile states is naturally achieved. We review the possibility of generating the observed baryon asymmetry of the universe via leptogenesis where the CP violation comes solely from the low-energy CP phases. In addition, we study the effect of texture zero in the Dirac mass matrix on the parameter space of the model for successful resonant leptogenesis.}

\begin{document}
\maketitle
\flushbottom

\section{Introduction}
\label{sec:intro}
Despite the success of the Standard Model (SM) of particle physics, it fails to explain the observed tiny neutrino masses and the observed baryon asymmetry of the Universe (BAU). The remedy to such a limitation is the extension of the SM with extra sterile singlets like the seesaw models \cite{minkowski1977mu, mohapatra1980neutrino, yanagida1980horizontal, gell2010complex, glashow1980future} The simplest extension of the SM is the type-I seesaw mechanism, where the observed BAU can be explained via thermal leptogenesis \cite{fukugita1986barygenesis}. In the type-I seesaw mechanism, the SM is extended by adding singlet right-handed (RH) neutrinos, and their out-of-equilibrium, lepton number violating decays produce the lepton asymmetries, which are then processed in baryon asymmetries via sphaleron interactions \cite{kuzmin1985anomalous}.

In the standard type-I seesaw with hierarchical RH neutrino mass spectrum, successful leptogenesis introduces a lower bound on the mass scales of the RH neutrino of about $10^9$ GeV \cite{davidson2002lower}. However, the mass scale for successful leptogenesis may be lowered in a scenario where the RH neutrinos have a quasi-degenerate mass spectrum.

Another theoretically motivated extension of SM with low-scale sterile neutrinos is the inverse seesaw (ISS) mechanism, where two types of sterile neutrinos: RH neutrinos and SM gauge singlets, are introduced to the SM \cite{gonzalez1990isosinglet, deppisch2005enhanced, abada2014looking, arina2008minimal, dev2010tev,mukherjee2023rescuing}. In this mechanism, the light neutrino mass is doubly suppressed allowing the Yukawa coupling to be much larger at the TeV scale of sterile neutrino mass. The quasi-degeneracy in the sterile neutrinos can be naturally realized due to the presence of a small lepton-number violating mass parameter $\mu$ \cite{Wyler:1982dd, mohapatra1986mechanism, mohapatra1986neutrino, ma1987lepton, gonzalez1989fast, gonzalez1990isosinglet, deppisch2005enhanced}. As the mass parameter $\mu$ allows the sterile states to acquire a quasi-degenerate mass scale, resonantly enhanced leptogenesis can be naturally realized in such a model without fine-tuning. The ISS mechanism is a low-scale model making it testable in future experiments. Since the mass scale of the sterile neutrinos relevant to leptogenesis is low, one needs to consider a fully-flavored leptogenesis regime. In a low-scale leptogenesis, the low-energy CP-violating phases can related to the high-energy CP-violation necessary to produce the BAU.

A similar study on resonant leptogenesis within the context of the ISS(2,2) model has been previously considered in Ref. \cite{Chakraborty_2022}. In this study, the authors have analysed resonant leptogenesis considering different specific regions for the mass of the lightest sterile state. Their study also includes constraints from the Lepton Flavour Violating (LFV) processes. 

Motivated by the fact that the measurements of the low-energy CP-phases are not as precise as the neutrino mixing angles in the neutrino oscillation experiments, we aim to study the effect of these phases on leptogenesis. We consider a minimal form of ISS mechanism that can accommodate the neutrino oscillation experimental data i.e., SM extended with two RH neutrinos and two sterile gauge singlets \cite{abada2014looking, malinsky2009non, hirsch2010minimal, blanchet2010leptogenesis, dias2012simple, agashe2019natural, Abada_2017, Fern_ndez_Mart_nez_2023}. We then proceed to find the parameter space of the defined model such that the BAU is generated effectively.

This paper is organized as follows. In section \ref{sec:1}, we describe the framework of the ISS(2, 2) model and define the three scenarios on which we study the implications of leptogenesis. In section \ref{sec:2}, we review how resonant leptogenesis can be naturally achieved within the ISS(2, 2) framework. It also includes the results of our numerical analysis beginning with section \ref{sec:2a} where we show the results of the parameter scan for the scenario where the $R$ matrix has complex entries. Section \ref{sec:2b} presents the results of the analysis for the case where $R$ has a special form with real entries and the parameter space for the third scenario with texture zeros in Dirac mass matrix is presented in section \ref{sec:2c}. We finally summarize our conclusions in section \ref{sec:con}.

\section{Model Framework}
\label{sec:1}
We have considered the extension of SM by introducing two RH neutrinos and two SM gauge singlet fermions which results in a minimal form of ISS and is denoted by ISS(2,2). The Lagrangian of the model which is invariant under the SM gauge symmetry is,
\begin{equation}
\label{eq:1}
	-\mathcal{L}_\nu = Y_\nu \bar{l}_L \tilde{H} N_R + M_R \bar{\left(N_R\right)^c} \left(S_L\right)^c + \frac{1}{2} \mu \bar{S_L} \left(S_L\right)^c + h.c.,
\end{equation}

where $\bar{l}_L$ is the SM lepton doublet, $\tilde{H} = i \sigma_2H^*$ with $H$ being the SM Higgs doublet and $\sigma$ denotes the $2^{nd}$ Pauli matrix. The extension of the SM includes the RH neutrinos $N_R$ and the SM gauge singlets $S_L$. As the Higgs doublet $H$ acquires vacuum expectation value ($vev$) and the gauge symmetry is broken i.e., $SU(2)_L \otimes U(1)_Y \rightarrow U(1)_{EM}$, we obtain the light neutrino mass matrix,
\begin{equation}
\label{eq:2}
	M_\nu =
		\begin{pmatrix}
			0     & m_D   & 0   \\
			m_D^T & 0     & M_R \\
			0     & M_R^T & \mu
		\end{pmatrix},
\end{equation}

where $m_D = Y_\nu/\sqrt{2}$ represents the Dirac mass matrix, $M_R$ is a complex $2 \times 2$ mass matrix and $\mu$ is a complex, symmetric $2 \times 2$ matrix. With $\mu \ll m_D \ll M_R$, diagonalization of equation \ref{eq:2} gives,

\begin{equation}
\label{eq:3}
	m_\nu = m_D \left(M_R^T\right)^{-1} \mu \left(M_R\right)^{-1} m_D^T .
\end{equation}

As can be seen in equation \ref{eq:3}, the light neutrinos are suppressed by the smallness of $\left(m_D M_R^{-1}\right)^2$ as well as the parameter $\mu$. In the one-generation case, the  masses of the light neutrinos can be reproduced for the following scale of the different mass states \cite{deppisch2005enhanced},
\begin{equation}
\label{eq:4}
	\left(\frac{m_\nu}{0.1 ~\textrm{eV}}\right) = \left(\frac{m_D}{100 ~\textrm{GeV}}\right)^2 \left(\frac{\mu}{1 ~\textrm{keV}}\right) \left(\frac{M_R}{10^4~ \textrm{GeV}}\right)^{-2},
\end{equation}

The neutrino mass matrix of equation \ref{eq:3} can be approximately diagonalized by the unitary matrix $U_{\textrm{PMNS}}$ as follows,
\begin{equation}
\label{eq:5}
	U^\dagger_{\textrm{PMNS}} m_\nu U^*_{\textrm{PMNS}} = \textrm{diag}(m_1,m_2,m_3) = m_d
\end{equation}

where the unitary matrix in the standard parametrization is represented by,
\begin{equation}
\label{eq:6}
    U_{\textrm{PMNS}} = 
    \begin{pmatrix}
        c_{12} c_{13} & s_{12} c_{13} & s_{13} e^{-i \delta}\\
        -s_{12} c_{23} - c_{12} s_{23} s_{13} e^{i \delta} & c_{12} c_{23} - s_{12} s_{23} s_{13} e^{i \delta} & s_{23} c_{13}\\
        s_{12} s_{23} - c_{12} c_{23} s_{13} e^{i \delta} & -c_{12} s_{23} - s_{12} c_{23} s_{13} e^{i \delta} & c_{23} c_{13}
    \end{pmatrix}
    \times
    \begin{pmatrix}
        1 & 0 & 0\\
        0 & e^{i \alpha_{21}/2} & 0\\
        0 & 0 & e^{i \alpha_{31}/2}
    \end{pmatrix}
\end{equation}

A salient feature of the ISS(2,2) model is that one of the three light neutrinos is massless and thus one of the Majorana CP phases becomes unphysical. To be more specific, for the case of normal hierarchy (NH) we have $m_1=0 < m_2 < m_3$ with the single Majorana phase redefined as $\sigma = (\alpha_{21}-\alpha_{31})/2$  and $m_3=0 < m_1 < m_2$ for inverted hierarchy (IH) with the single Majorana phase given as $\sigma = \alpha_{21}/2$.

Using equations \ref{eq:3} and \ref{eq:5}, we can derive the Casas-Ibarra \cite{casas2001oscillating} type parametrization of the Dirac mass matrix for the inverse seesaw model as \cite{dolan2018dirac},
\begin{equation}
    \label{eq:7}
    \left(m_d^{-1/2} U_{\textrm{PMNS}}^\dagger m_D \left(M_R^T\right)^{-1} \mu^{1/2}\right)\cdot \left(\mu^{1/2} M_R^{-1} m_D^T U_{\textrm{PMNS}}^*   m_d^{-1/2}\right) = \mathbb{I},
\end{equation}
which leads to,
\begin{equation}
    \label{eq:8}
    m_D = U_{\textrm{PMNS}}~m_d^{1/2}~R~\mu^{-1/2}~M_R^T
\end{equation}
where R  is a complex  $3\times 2$ matrix given by,

\begin{align}
    \label{eq:9}
    R &= \begin{pmatrix}
        0 & 0 \\
        \cos{\zeta} & -\sin{\zeta} \\
        \sin{\zeta} & \cos{\zeta}
    \end{pmatrix} ~~~~~~~~~~ \textrm{for NH}\\
    R &= \begin{pmatrix}
        \cos{\zeta} & -\sin{\zeta} \\
        \sin{\zeta} & \cos{\zeta} \\
        0 & 0
    \end{pmatrix} ~~~~~~~~~~ \textrm{for IH}
\end{align}
with $z = Re(\zeta)+ i~Im(\zeta)$, in general being a complex parameter. In the next section of the paper, we analyze the parameter space for successful resonant leptogenesis for three different scenarios in the ISS(2, 2) model, namely, ($i$) $R$ matrix with complex $\zeta$: here the source of CP-violation comes from the matrix $R$ and the CP-phases present in the PMNS matrix, ($ii$) $R$ matrix with $Re(\zeta) = \pi/4$ and  $Im(\zeta) = 0$ : in such a special case CP-violation necessary for leptogenesis comes from the Dirac and Majorana CP-phases, and ($iii$) texture zeros in $m_D$: in this scenario we consider one of the elements of $m_D$ to be zero such that we can write the parameter $\zeta$  in terms of the neutrino mass, mixing angles and the CP-phases of the $U_{\textrm{PMNS}}$ matrix. Thus, just as in scenario ($ii$), the CP-phases present in the PMNS matrix provide the necessary CP-violation.

\section{Resonant leptogenesis}
\label{sec:2}
To study leptogenesis in the ISS(2,2) model, we work in the basis where the sterile neutrino mass sub-matrix is real and diagonal. The lower $2 \times 2$ block of the mass matrix presented in the equation \ref{eq:2} is given as 
\begin{equation}
\label{eq:11}
    M_{SN} = 
    \begin{pmatrix}
        0 & M_R \\
        M_R^T & \mu        
    \end{pmatrix}
\end{equation}
We perform a block diagonalization on the above matrix such that we are working in the basis where the sterile neutrinos are in their mass basis. This transforms the mass matrix given in the equation \ref{eq:2} by rotating the Yukawa couplings to the SM leptons and can be written as
\begin{equation}
    \label{eq:12}
    M_\nu \rightarrow M_\nu \simeq 
    \begin{pmatrix}
        0 & m_D^{'} & 0 \\
        \left(m_D^{'}\right)^T & M_R - \frac{1}{2}\mu & 0 \\
        0 & 0 & M_R + \frac{1}{2}\mu
    \end{pmatrix},
\end{equation}
where $m_D^{'}$ is the rotated Yukawa coupling matrix. It is clear from the equation \ref{eq:12} that for small values of $\mu$ the mass spectrum of the heavy sterile neutrinos becomes degenerate and the scenario of resonant leptogenesis can be naturally achieved. In resonant leptogenesis, the CP-violating, out-of-equilibrium decay of the degenerate sterile neutrinos produces the observed BAU. A non-zero lepton asymmetry is generated from the CP-violation obtained from the interference of the tree-level decay of the sterile neutrinos with the one-loop level. In the case of degenerate sterile neutrinos, the self-energy correction is resonantly enhanced and the flavor-dependent CP asymmetry is defined as \cite{pilaftsis1997cp, anisimov2006cp, de2007resonant}
\begin{equation}
    \label{eq:13}
    \varepsilon_i^\alpha = \frac{\Gamma \left(N_{i\alpha}\rightarrow l_\alpha \Phi\right) - \Gamma \left(N_{i\alpha}\rightarrow l_\alpha^c \Phi^\dagger\right)}{\sum_\alpha \left[\Gamma \left(N_{i\alpha}\rightarrow l_\alpha \Phi\right) + \Gamma \left(N_{i\alpha}\rightarrow l_\alpha^c \Phi^\dagger\right)\right]}
\end{equation}
The CP-asymmetry parameters are given by
\begin{equation}
    \label{eq:14}
    \varepsilon^\alpha_i=\sum_{i \neq j}\frac{\textrm{Im}\left[h_{i\alpha}^\dagger h_{\alpha j}\left(h^\dagger h\right)_{ij}\right] + \frac{M_i}{M_j}\textrm{Im}\left[h_{i\alpha}^\dagger h_{\alpha j}\left(h^\dagger h\right)_{ji}\right]}{\left(h^\dagger h\right)_{ii} \left(h^\dagger h\right)_{jj}}\cdot \frac{\left(M_i^2 - M_j^2\right)\cdot M_i \Gamma_j}{\left(M_i^2 - M_j^2\right)^2 + M_i^2 \Gamma_j^2}.
\end{equation}
where, $h = \frac{\sqrt{2}}{v}m_D^{'}$,  $\Gamma_i$ is the decay width of the heavy sterile neutrino state $N_i$, and $M_i$ is the mass eigenvalue of $N_i$. A lepton asymmetry is generated by utilizing the above CP-asymmetry parameter by solving the Boltzmann equation. The Boltzmann equation under consideration is a coupled differential equation describing the time evolution of the density of  heavy sterile neutrinos, $n_{N_i}$, and the lepton number density $n_{N_{\alpha \alpha}}$ (with $\alpha = e, \mu, \tau$) \cite{de2007quantum},

\begin{align}
    \label{eq:15}
    \frac{dn_{N_i}}{dz} &= -D_i \left(n_{N_i} - n_{N_i}^{eq}\right)\nonumber\\ 
    \frac{n_{N_{\alpha\alpha}}}{dz} &= \sum_{i=1}^2 \varepsilon_i^\alpha D_i \left(n_{N_i} - n_{N_i}^{eq}\right) - W_{ID}~n_{N_{\alpha\alpha}}
\end{align}

where 
\begin{align}
    \label{eq:16}
    W_{ID} &= \frac{1}{4} K z^3 \mathcal{K}_1(z),\\
    \label{eq:17}
    D_i &= \frac{z}{H(z=1)}\cdot\frac{\Gamma_{i}}{n_{N_{i}}^{eq}},
\end{align}

denotes the washout due to inverse decay and the decay term,  respectively, with $\mathcal{K}_1(z)$  being the modified Bessel function of the first kind, $K=\tilde{m}/m_*$ is known as the decay parameter, $n_{N_i}^{eq}$ is the equilibrium number density of $N_i$ and is defined as
\begin{equation}
    \label{eq:18}
    n^{eq}_{N_i} = \frac{3}{8} z^2 \mathcal{K}_2(z) 
\end{equation}
The parameter $H$ is the Hubble parameter, $\tilde{m}$  is the effective neutrino mass, and $m_*$ is the equilibrium neutrino mass.

The baryon asymmetry, $\eta_B$ can be estimated by solving equations \ref{eq:15}, and the value of $\eta_B$ depends on the initial condition of the heavy sterile neutrinos. The numerical evaluation of baryon asymmetry performed in this work involves the decay of sterile neutrinos which is a good approximation, and we do not consider other effects such as the scattering process, spectator effects, thermal corrections, etc.  In our analysis, we choose $M_R$ and $\mu$ matrices to be diagonal with a degenerate mass spectrum of right-handed neutrinos, $M_{R_1} = M_{R_2} = 1$ TeV. The elements of the matrix $\mu$ determine the level of degeneracy among the heavy sterile states and we take $\mu_{i}$
 to be free parameters that will be constrained by successful leptogenesis. Further, we choose a vanishing initial abundance of sterile neutrinos and scan over the parameters of our model, namely:  $\delta,~\sigma,~\mu_{1},~\mu_{2},~Re(\zeta),~Im(\zeta)$.  We make a parameter scan by using flat prior and define $1\sigma$, $2\sigma$, and $3\sigma$ regions of agreement with the observed value of baryon asymmetry by evaluating the log-likelihood function at a point, $p$. The function is given as 
 \begin{equation}
     \label{eq:19}
     log~L = -\frac{1}{2}\left(\frac{\eta_B^2(p) - \eta_{B_{CMB}}^2}{\Delta \eta_{B_{CMB}}^2}\right)
 \end{equation}
Depending on the scenarios discussed in section \ref{sec:1}, we will have different dimensions of  $p$. 

\subsection{Scenario I: $R$ matrix with complex $\zeta$}
\label{sec:2a}
\begin{figure}[tbp]
\centering 
\includegraphics[width=.3\textwidth]{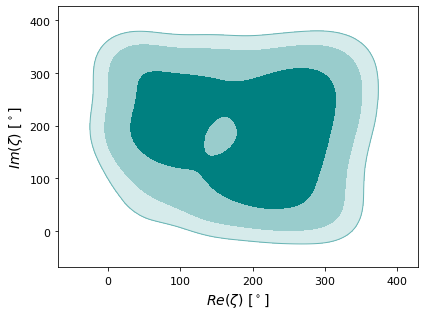}
\hfill
\includegraphics[width=.3\textwidth]{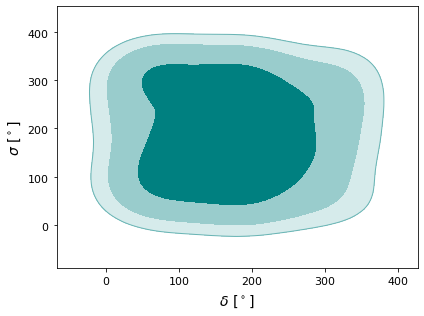}
\hfill
\includegraphics[width=.3\textwidth]{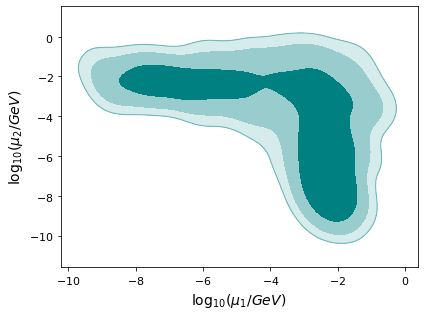}

\medskip

\includegraphics[width=.3\textwidth]{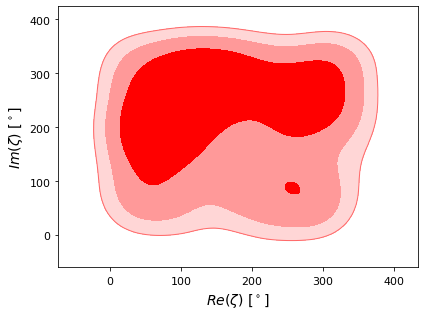}
\hfill
\includegraphics[width=.3\textwidth]{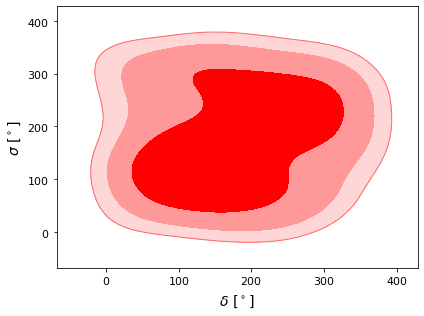}
\hfill
\includegraphics[width=.3\textwidth]{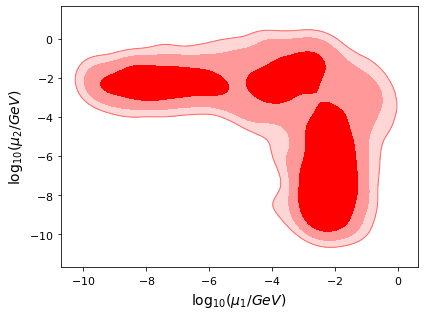}
\caption{The projection for leptogenesis with the contours representing the $1\sigma$, $2\sigma$, and $3\sigma$ confidence levels. The teal and red colour denote the case of NH and IH, respectively.}
\label{fig:1}
\end{figure}

This is the most general case with $p = (\delta,~\sigma,~\mu_{1},~\mu_{2},~Re(\zeta),~Im(\zeta))$, being the point in the 6-dimensional parameter space of the model over which the scan is made using the \texttt{MultiNest} package \cite{feroz2009multinest}. The results of our analysis are shown in figure \ref{fig:1}. It shows the 2-dimensional projection for leptogenesis with CP violation coming from the complex matrix $R$ as well as the phases of the PMNS matrix. The mass splitting between the two sterile states, which is necessary for successful resonant leptogenesis is quantified by the parameters $\mu_1$, and $\mu_2$. Figure \ref{eq:1} also demonstrates the variation of baryon asymmetry $\eta_B$ in the $(\mu_1 - \mu_2)$ space. The best-fit values for the parameters of the model are $\delta = 194^\circ$, $\sigma = 152^\circ$, $\log_{10}(\mu_1/\textrm{GeV}) = -3.6$, $\log_{10}(\mu_2/\textrm{GeV}) = -3.0$, $Re(\zeta) = 154^\circ$, $Im(\zeta) = 193^\circ$ in the NH case. Clearly, the measured value of $\delta$ coincides with the global fit of experimental data presented in NuFit 5.1 \cite{esteban2020fate} for the NH case. For the IH case, the best-fit values are $\delta = 203^\circ$, $\sigma = 216^\circ$, $\log_{10}(\mu_1/\textrm{GeV}) = -2.9$, $\log_{10}(\mu_2/\textrm{GeV}) = -2.8$, $Re(\zeta) = 175^\circ$, $Im(\zeta) = 212^\circ$. In this case, the measured best-fit value of $\delta$ lies outside the $1\sigma$ range of the global fit of the neutrino oscillations experimental data.  For the range of $\delta$ between $[92 - 296]^\circ$ we obtain the $1 \sigma$ value of the observed baryon asymmetry, $\eta_B$ in the NH case and $[117 - 289]^\circ$ in the IH case. 

\subsection{Scenario II: $R$ is a real matrix with $Re(\zeta) = \pi/4$ and $Im(\zeta) = 0$}
\label{sec:2b}
\begin{figure}[tbp]
\centering 
\includegraphics[width=.45\textwidth]{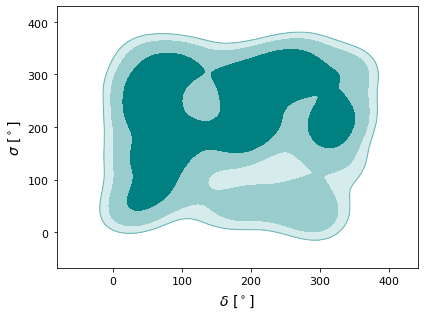}
\hfill
\includegraphics[width=.45\textwidth]{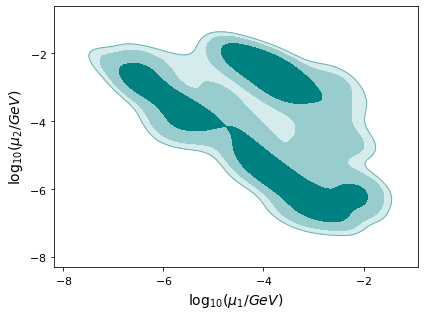}

\medskip

\includegraphics[width=.45\textwidth]{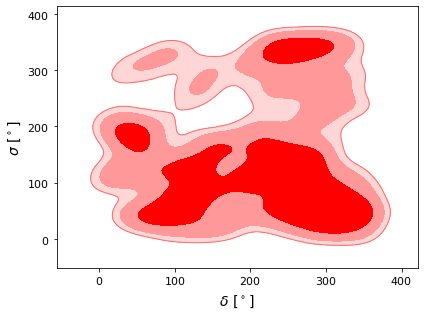}
\hfill
\includegraphics[width=.45\textwidth]{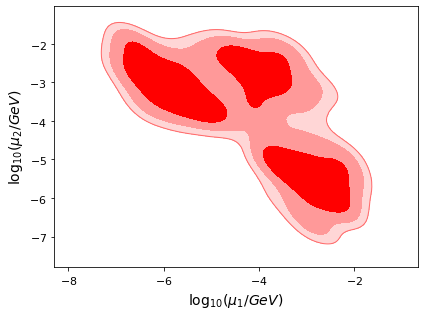}
\caption{The projection for leptogenesis with the contours representing the $1\sigma$, $2\sigma$, and $3\sigma$ confidence levels. The teal and red colour denotes the case of NH and IH, respectively.}
\label{fig:2}
\end{figure}

For the second case, we choose $Re(\zeta) = \pi/4$ and $Im(\zeta) = 0$ making the $R$ matrix real such that the necessary CP violation comes solely from the CP phases present in the neutrino mixing matrix. It is clear that the generation of BAU has a contribution from high as well as low energy parameters, however, in this case, we analyze the effect of low-energy CP phases on leptogenesis in the ISS(2,2) model. We first make a parameter scan in a 4-dimensional parameter space with $p = (\delta,~\sigma,~\mu_1,~\mu_2)$. Figure \ref{fig:2} shows the allowed region for the parameters of the model. The contours represent the region for which the observed value of baryon asymmetry can be obtained within $1\sigma$, $2\sigma$, and $3\sigma$ confidence intervals. The best-fit values for the parameters are $\delta = 167^\circ$, $\sigma = 216^\circ$, $\log_{10}(\mu_1/\textrm{GeV}) = -4.0$, $\log_{10}(\mu_2/\textrm{GeV}) = -3.9$ in the NH case. For the IH case, the best-fit values are $\delta = 241^\circ$, $\sigma = 109^\circ$, $\log_{10}(\mu_1/\textrm{GeV}) = -4.2$, $\log_{10}(\mu_2/\textrm{GeV}) = -3.5$. For the range of $\delta$ between $[46 - 288]^\circ$ we obtain the $1 \sigma$ value of the observed baryon asymmetry, $\eta_B$ in the NH case and $[140 - 342]^\circ$ in the IH case. We find that the value of $\delta$ calculated within the model that obtains the best-fit value to the observed BAU lies within the $1\sigma$ region of NuFit 5.1 data for the case of NH, however, it lies slightly outside the $1\sigma$ range for the IH case.  

\subsection{Scenario III: texture zeros in $m_D$}
\label{sec:2c}
We consider one of the elements of the Yukawa matrix, $m_D$ to be zero. Considering the (1, 1) element of $m_D$ to be zero we obtain a simple expression using Casas-Ibarra type parametrization of equation \ref{eq:8}
\begin{equation}
    \label{eq:20}
    \left(U_{\textrm{PMNS}}\right)_{21} \sqrt{m_2} \cos\zeta + \left(U_{\textrm{PMNS}}\right)_{23} \sqrt{m_3} \sin\zeta = 0,
\end{equation}
in the NH case, and,

\begin{equation}
    \label{eq:21}
    \left(U_{\textrm{PMNS}}\right)_{11} \sqrt{m_1} \cos\zeta + \left(U_{\textrm{PMNS}}\right)_{22} \sqrt{m_2} \sin\zeta = 0,
\end{equation}
in the IH case.

\begin{figure}[tbp]
\centering 
\includegraphics[width=.45\textwidth]{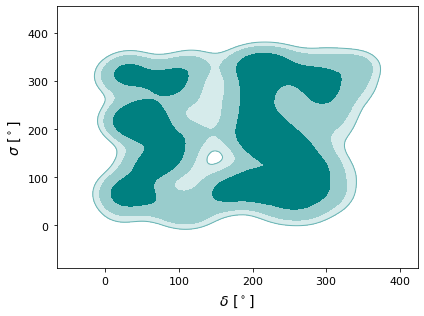}
\hfill
\includegraphics[width=.45\textwidth]{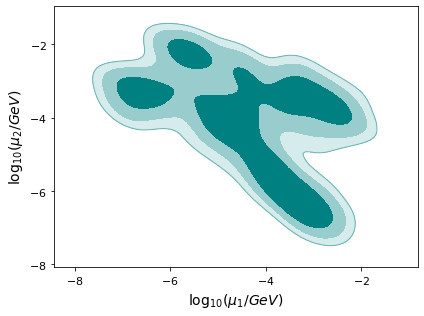}

\medskip

\includegraphics[width=.45\textwidth]{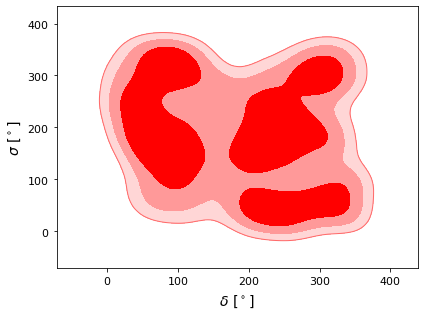}
\hfill
\includegraphics[width=.45\textwidth]{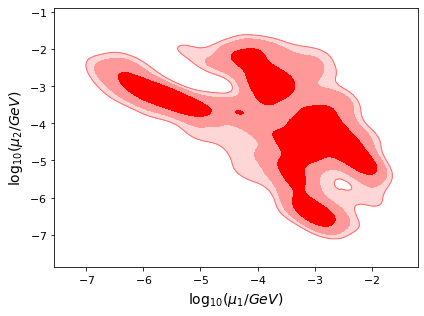}
\caption{The projection for leptogenesis with the contours representing the $1\sigma$, $2\sigma$, and $3\sigma$ confidence levels. The teal and red colour denotes the case of NH and IH, respectively.}
\label{fig:3}
\end{figure}

From the above relations, one can write the parameter $\zeta$ in terms of the neutrino masses, mixing angles, and the CP phases of the PMNS matrix. We have a 4-dimensional parameter space with a particular point defined as $p =(\delta,~\sigma,~\mu_{1},~\mu_{2})$. The results of the exploration are presented in figure \ref{fig:3}. The contours represent the region of the parameter space for which the observed value of baryon asymmetry lies within $1\sigma$, $2\sigma$, and $3\sigma$ confidence intervals. For the parameters of the model the best-fit value are $\delta = 196^\circ$, $\sigma = 192^\circ$, $\log_{10}(\mu_1/\textrm{GeV}) = -4.1$, $\log_{10}(\mu_2/\textrm{GeV}) = -3.9$ in the NH case. For the IH case, the best-fit values are $\delta = 204^\circ$, $\sigma = 182^\circ$, $\log_{10}(\mu_1/\textrm{GeV}) = -3.9$, $\log_{10}(\mu_2/\textrm{GeV}) = -3.4$. For the range of $\delta$ between $[86 - 306]^\circ$ we obtain the $1 \sigma$ value of the observed baryon asymmetry, $\eta_B$ in the NH case and $[90 - 318]^\circ$ in the IH case. From the best-fit values of the Dirac CP phase $\delta$, we see that for both the NH as well as the IH cases the measured value agrees with the experimental data up to $1\sigma$ confidence level.

\section{Conclusion}
\label{sec:con}
In this paper, we have studied resonant leptogenesis in the framework of the minimal form of the inverse seesaw model ISS(2, 2). Here, the SM is extended by adding two right-handed and two SM gauge singlet neutrinos. Considering the quasi-degenerate,  quasi-Dirac sterile neutrino states, we study the scenario of resonant leptogenesis. 

To carry out our analysis, we write the Dirac mass matrix of the ISS(2, 2) model in the form of Casas-Ibarra type parametrization. We investigate the viable parameter space for leptogenesis in the ISS(2, 2) model. We explored the parameter space for three different scenarios. Firstly, we consider the case where the CP violation necessary for successful leptogenesis comes from both high-energy parameters (in the form of complex $R$ matrix) and low-energy CP phases. Secondly, we assess the possibility of CP violation arising exclusively from the low-energy leptonic sector in the form of CP phases present in the PMNS matrix. Finally, the third case involves texture zero in the Dirac mass matrix which allows us to write the elements of the complex $R$ matrix in terms of low-energy parameters. In other words, we explore the possibility of successful low-energy resonant leptogenesis with CP violation coming from the phases present in the PMNS matrix.

We take the best-fit values of the three mixing angles and two mass-squared differences as the input for the Casas-Ibarra parametrization and allow the complex parameter, $\zeta$, the elements of the matrix $\mu$ and the two CP phases present in the PMNS matrix to run freely in a particular region as free parameters. We numerically solve the coupled Boltzmann equation that describes the evolution of the Lepton asymmetry and eventually gives the baryon asymmetry. We make a parameter exploration for the free parameters of the model using the measured value of baryon asymmetry. From the measured value of $\delta$, we find that the model agrees with the experimental data of the Dirac CP phase up to $1\sigma$ confidence level in the NH case for all three scenarios, however, in the IH case it only agrees with the scenario where we consider texture zeros in the Dirac mass matrix. Future precision experiments may give much more stringent results on the Dirac CP phase and probe such a model.   
 
\section*{Acknowledgement}
BT acknowledges the Department of Science and Technology (DST), Government of India for INSPIRE Fellowship vide Grant No. DST/INSPIRE/2018/IF180588. The research of NKF is funded by DST-SERB, India under Grant No. EMR/2015/001683.

\bibliographystyle{JHEP}
\bibliography{references}

\end{document}